\documentclass[screen, nonacm, acmsmall, sigconf, numbers]{acmart}
\AtBeginDocument{%
  }

\usepackage{orcidlink}
\usepackage{algorithm}%
\usepackage{algorithmicx}%
\usepackage[noend]{algpseudocode}%

\setcopyright{none}

\usepackage{caption}
\makeatletter
\renewcommand{\section}{%
  \@startsection{section}{1}{\z@}%
    {-2.0ex \@plus -0.5ex \@minus -0.2ex}%
    {1.0ex \@plus 0.2ex}%
    {\LARGE\bfseries}}   
\makeatother
\makeatletter
\renewcommand{\subsection}{%
  \@startsection{subsection}{2}{\z@}%
    {-1.5ex \@plus -0.5ex \@minus -0.2ex}%
    {0.8ex \@plus 0.2ex}%
    {\Large\bfseries}}  
\makeatother

\begin{document}

\title{SPHaptics: A Real-Time Bidirectional Haptic Interaction Framework for Coupled Rigid--Soft Body and Lagrangian Fluid Simulation in Virtual Environments}


\author{\orcidlink{0009-0005-2896-7082} William Baumgartner}
\authornote{Corresponding author. E-mail: wnb2013@nyu.edu}
\email{wnb2013@nyu.edu}
\affiliation{%
  \institution{New York University}
  \city{New York}
  \state{NY}
  \country{USA}
}

\author{\orcidlink{0000-0002-7811-9357} Gizem Kayar-Ceylan}
\email{gk2409@nyu.edu}
\affiliation{%
  \institution{New York University}
  \city{New York}
  \state{NY}
  \country{USA}
}

\renewcommand{\shortauthors}{Baumgartner and Kayar-Ceylan}

\begin{abstract}
  Haptic feedback enhances immersion in virtual environments by allowing users to physically interact with simulated objects. Supporting accurate force responses in multiphysics systems is challenging because physically based simulation of fluid, rigid, and deformable materials is computationally demanding, especially when interaction must occur in real time. We present a unified framework for real-time, bidirectional haptic interaction with rigid bodies, deformable objects, and Lagrangian fluids in virtual reality (VR). Our approach integrates Smoothed Particle Hydrodynamics (SPH) with two-way force coupling and feedback smoothing to maintain stability and produce physically meaningful tactile responses. This enables users to manipulate objects immersed in fluid and feel reaction forces consistent with fluid-structure behavior. We demonstrate the capabilities of our framework through interactive VR scenarios involving fluid stirring, soft tissue manipulation, and rigid-body interaction. The proposed system advances haptic-enabled multiphysics simulation by unifying fluid, soft-body, and rigid-body dynamics into a single platform suitable for immersive educational applications.
\end{abstract}

\keywords{Smoothed Particle Hydrodynamics, Haptics, Real-Time Simulation, Virtual Reality, Education}
\begin{teaserfigure}
  \includegraphics[width=\textwidth]{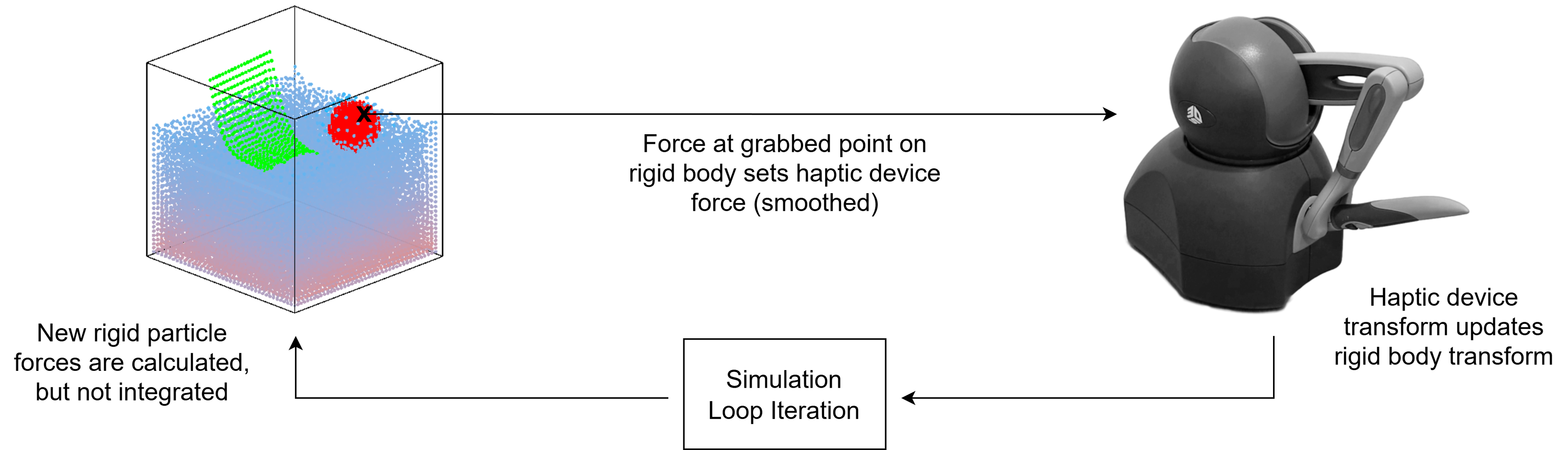}
  \caption{The haptic feedback loop. Force at the grabbed point on the rigid body sets the haptic device force, and the device's motion updates the rigid body transform for the next simulation iteration.}
  \label{fig:diagram}
\end{teaserfigure}

\maketitle

\section{Introduction}\label{sec1}

Haptic interaction has emerged as one of the key components of immersive and physically engaging virtual experiences. By providing force feedback that aligns with the user's tactile perception, haptics bridges the sensory gap between the real and virtual worlds, enabling direct interaction with simulated objects through both motion and touch \cite{Emami2025}. When integrated with physics-based simulations, haptic feedback can convey material characteristics and dynamic behavior, improving realism and helping users better understand complex physical phenomena. However, supporting stable and accurate haptic rendering in multiphysics environments remains challenging due to the computational cost of physical simulation, the need for responsiveness, and the strict stability requirements of high-frequency force-feedback loops.

Previous haptic simulation systems have largely focused on exclusively rigid/soft body or fluid interactions, avoiding the complexities of a coupled simulation \cite{Baxter2005, Mora2008}. Yet, many real-world phenomena, e.g. object immersion, drag, buoyancy, and deformation under flow, are inherently governed by the coupled dynamics of solids and fluids. Realistic simulation of these interactions requires consistent two-way coupling, in which the motion and forces of solid bodies influence the fluid flow, and vice versa. For virtual reality (VR) applications, this coupling must occur at real-time rates to maintain physical accuracy and responsiveness. Achieving this bidirectional interaction in real time remains a major challenge, since accurate fluid dynamics are computationally expensive while haptic feedback requires fast and stable force updates.

To address this, our work introduces a unified framework for bidirectional haptic interaction between rigid and deformable solids and Lagrangian fluid models, implemented within a real-time VR physics platform. Through this architecture, users can directly manipulate virtual objects immersed in fluid, observe physically accurate responses, and feel reactive forces in real time. The system thus extends the boundaries of touch-based simulation, enabling physically coherent interactive experiences that merge visual immersion with tactile realism.

A key computational component of this framework is smoothed particle hydrodynamics (SPH), a mesh-free particle-based method widely used to simulate fluids and deformable continua \cite{Monaghan1999}. SPH represents fluids as a set of discrete particles that carry mass, momentum, and thermodynamic properties, with interactions governed by smoothing kernels that approximate continuum equations. Its Lagrangian nature makes SPH particularly well suited for handling free-surface flows, fluid--solid coupling, and topological changes, all of which are challenging for grid-based methods. Despite these advantages, the integration of SPH into real-time VR and haptic systems remains relatively limited. The primary barriers include the high computational cost, the difficulty in maintaining stability under large time-step constraints, and the challenge of aligning the continuous high-frequency haptic control loop with discrete physics simulation updates. Consequently, only a small number of VR applications have leveraged SPH for interactive, touch-enabled fluid dynamics, and even fewer have addressed full two-way coupling with deformable solids \cite{Cirio2011}.

\textbf{Our contribution.} In this paper, we present an approach that bridges these gaps by optimizing SPH-based fluid--deformable/rigid coupling for real-time, haptic-enabled VR environments. Our contributions include: (1) a stable coupling scheme for two-way interaction between rigid/soft solids and Lagrangian fluids; (2) a haptic feedback loop control system integrated within a unified multiphysics simulation; and (3) a prototype VR system that demonstrates physically consistent, tactilely perceivable interactions between virtual fluids and solids. The proposed framework advances the integration of SPH-based multiphysics simulation with immersive, touch-driven virtual experiences.

\section{Related work}\label{sec2}

In this section, we will review and discuss previous research that is relevant to this project. We will briefly go over the current state of SPH techniques and their applications, especially noting extensions beyond fluids. Additionally, we will discuss VR and its uses for educational applications that this work draws on. We also review the state of haptic technology, specifically the application of motion-enabled devices to virtual simulations similar to the one we propose.

\subsection{SPH}\label{subsec2.1}

SPH has become a widely used technique for fluid and continuum simulation in computer graphics and physically based modeling. Comprehensive surveys of SPH methods in computer graphics and scientific computing describe steady advances in numerical stability, kernel functions, and boundary handling \cite{Xi2020}. Recent overviews also highlight the growing shift toward GPU acceleration and real-time simulation, driven by the needs of interactive visual and virtual environments. Extensions for real-time engines such as Unity demonstrate efficient particle management and kernel optimization techniques capable of maintaining fluid continuity and volume preservation at high frame rates \cite{Sun2025}. SPH has been shown to be practical to many fields, including astrophysics, environmental science, and biomedical simulation \cite{Springel2010, Cleary2004, Muller2004}. These developments establish SPH as a versatile and computationally scalable framework for physically based fluid simulation in immersive applications.

Beyond fluids, recent research has also extended SPH toward representations that couple fluids with rigid and deformable solids within a common particle-based framework. Unified approaches model both solid and fluid phases through the same SPH interaction laws, enabling two-way coupling through shared pressure and viscosity terms while enforcing distinct material constraints for elasticity and rigidity \cite{Koschier2019}. Earlier techniques demonstrated stable incompressible fluid-rigid coupling by treating solid surfaces as boundary particles and enforcing momentum exchange across interfaces \cite{Akinci2012}. More recent work on fluid-soft body coupling extends this principle by embedding elastic deformation models directly into the SPH domain, allowing accurate and stable interaction between fluid and deformable media \cite{AbuRumman2019}. Rigid and soft body integration into SPH simulations has been shown to be useful in robotics, materials science, and biomechanics \cite{Angelidis2025, Feng2024, Xu2023}.

Thus, current research shows the potential that SPH has as a unified particle-based framework for multiphysics simulation that captures fluid-structure interactions with accuracy. Despite its strengths, achieving real-time performance and stability for such coupled systems on consumer hardware is a difficult challenge, especially under haptic and VR constraints, and is the main motivation for this research.

\subsection{Virtual reality}\label{subsec2.2}

Virtual reality (VR) has rapidly evolved into a mainstream interactive technology that is capable of immersing users in computer-generated environments. Modern VR systems typically combine a form of head-mounted display (HMD), motion tracking, and haptic feedback to replace a user's perception with a simulated experience. Advancements in consumer-grade hardware such as improved displays, reduced latency, and input devices have significantly increased accessibility and usability. As a result, VR has transitioned from being used as a specialized research tool to now being a widely adopted platform across entertainment, education, and professional applications. VR enables intuitive interaction with virtual simulations where physical space or safety constraints might otherwise limit a real-world parallel \cite{Anthes2016}.

Moreover, educational virtual reality applications have become an increasingly viable approach for teaching people practical skills that translate to the real world. Unlike real-world education, VR can reduce constraints such as travel, material setup, and instructor availability, enabling access to consistent practice scenarios. Immersive simulations allow learners to practice complex or hazardous tasks safely in an environment such that they can receive guidance and feedback with no risks involved. This approach has been effectively demonstrated in recent work in medical areas, specifically in immersive visualization of brain lesions, exploration of human anatomy, and multi-sensory environments for individuals with neuro-developmental disorders \cite{Kelley2024, Izard2017, Yi2025}. Furthermore, VR has been used to simulate complicated lab setups in order to complement their real-life parallels, specifically in fields like chemistry and biology, where educational setups can often be infeasible or expensive \cite{Georgiou2007, Shim2003}.

Consequently, VR education is being adopted across a wide range of areas, including industrial operations, medical procedures, emergency response, and public safety, where authentic practice and performance improvement are critical \cite{Xie2021}. However, despite this progress, research specifically combining VR and fluid simulation remains relatively limited. Real-time fluid dynamics in VR environments introduces computational and perceptual challenges, specifically in balancing simulation accuracy with the low-latency required for comfortable VR experiences. Only a few studies have addressed these issues directly, such as that of Zhang et al. \cite{Zhang2019}, who proposed a GPU-based system for fast fluid animation in VR, and Wang et al. \cite{Wang2025}, who developed an adaptive fluid simulation method to improve VR rendering performance. These efforts show the emerging, yet still scarce focus on fluid interaction within VR, emphasizing the need for further research in this area that we hope to expand upon. 

\subsection{Haptics}\label{subsec2.3}

Haptic feedback enables users to physically interact with virtual environments by providing force and motion cues during manipulation of various objects. Haptic devices combine sensing of user input with force output to simulate contact with virtual objects or tools. This feedback can convey properties like weight, stiffness, and resistance, allowing users to feel the results of their actions, rather than only seeing them. Effective haptic interaction interfaces must accurately measure motion and apply responsive force output, which requires reliable sensing, fast actuation, and stable control of the feedback loop \cite{Emami2025}. Since we chose a specific haptic device for this research, the following techniques mainly focus on handling these challenges through software implementations.

In this research, we utilize the 3D Systems Touch haptic device \cite{3DSystems} (previously PHANToM Omni), which has 6 degrees of freedom (6-DoF), meaning it can translate and rotate in three dimensions. It can provide translational force with three degrees of freedom, allowing full integration with 3D virtual environments and interaction with virtual objects. The device measures position and velocity using encoders and applies torques to generate reaction forces to the user during contact \cite{Silva2009}. A key consideration when integrating such a device into physics based interactions is maintaining haptic stability, as force feedback must update at a high frequency. Sudden changes in the simulated forces or communication latency can lead to oscillations or unstable behavior, making control filtering and careful force rendering necessary to ensure smooth interaction. These challenges motivate approaches that balance real-time responsiveness with physically meaningful force computation, which we attempt to expand upon in this research.

The coupling of haptic interactions with accurate physical motion of rigid bodies has been attempted by several methods. A common strategy is to use a proxy or god-object model, where a virtual representation of the haptic device is constrained to the surface geometry while the real device moves freely. Forces are then computed by the difference between the proxy object and the device. This was originally developed for 3D point contact, but has been extended to 6-DoF rigid-body interactions \cite{Ortega2007, Zilles1995}. Other techniques introduce collision based force responses, such as applying impulse forces when rigid bodies collide, improving the accuracy of perceived forces without feeling overdamped \cite{Constantinescu2004}. This research utilizes ideas from both methods and focuses on real-time force integration that is smoothed to ensure stability.

Additionally, work has been done on handling soft (deformable) body haptic interactions, but it is a more complex problem to achieve a stable and accurate solution due to the many dynamic components. Most soft-body simulations use physically based models such as mass--spring systems or the finite element method (FEM) to provide deformations under applied forces \cite{Hui2006, DeSantis2024}. This research expands on such mass--spring systems by assigning SPH particles to their junctions to create a unified simulation.

Haptic feedback is widely used in virtual reality applications to increase immersion and support more realistic interaction with virtual objects. Devices range from grounded arm type systems, such as the device we experimented with, to handheld controllers and more recent wearable solutions such as haptic gloves \cite{Frisoli2024}. However, providing convincing haptic feedback in VR remains challenging due to limitations in actuation strength, device size, and hand tracking. Additionally, maintaining low latency and stable force output is difficult, particularly when simulating continuous interactions such as grasping, sliding, or lifting virtual objects \cite{Wee2021}. Thus, current VR haptic implementations often have to strike a balance between feedback realism, device comfort, and responsiveness. 

Haptic interaction with fluids also poses challenges due to the deformable and highly dynamic nature of fluid motion, requiring stable and responsive force computation, especially during real-time interaction \cite{Huang2025}. Early attempts at resolving haptic and fluid coupling relied on grid-based or simplified fluid models to approximate haptic forces. Eulerian approaches, for example, enabled interaction with paint or stirring effects but were limited by simple tool geometry and constrained fluid domains, reducing realism and flexibility \cite{Baxter2005, Mora2008}. More recent work uses smoothed-particle hydrodynamics (SPH) to better support free-surface flow and two-way coupling with arbitrarily shaped rigid bodies \cite{Liang2016}. Cirio et al. demonstrated 6-DoF force and torque feedback from SPH fluids, enabling stable interactions such as scooping or shaking fluid within containers at interactive rates \cite{Cirio2011}. Extending such techniques to also support rigid and soft bodies and more complex multiphysics couplings remains an open challenge which our work aims to address.

\section{Methods}\label{sec3}

Our goals for this research were to develop and evaluate a simulation framework that operates efficiently and with high accuracy. In order to accomplish this, there were many technical challenges regarding the real-time nature of the problem, coupled interactions, and controlling haptic feedback. In this section, we will discuss the problems we addressed and the solutions we propose.

\subsection{Real-time SPH}\label{subsec3.1}

At the core of our simulation is Smoothed-particle hydrodynamics (SPH), a Lagrangian (point-based) technique that enables us to simulate accurate fluid flow and interactions in real-time. SPH works by calculating the force on each particle at every timestep, and integrating their velocities and positions forward.

We calculate each particle's force by iterating over every particle and their neighboring particles to calculate the pressure force and viscosity force, as well as applying any external forces. The SPH algorithm performs multiple all-particle iterations and is described in Algorithm \ref{alg:sph}.
\begin{algorithm}
\caption{SPH Algorithm}\label{alg:sph}
\begin{algorithmic}
\ForAll{\text{$n$ particles $i$}}
    \ForAll{\text{$m$ neighbors $j$}}
        \State Calculate density, $\rho_{i}$ (\ref{density})
        \State Calculate pressure, $p_i$ (\ref{pressure})
    \EndFor
\EndFor
\ForAll{\text{$n$ particles $i$}}
    \ForAll{\text{$m$ neighbors $j$}}
        \State Calculate $F^{\text{pressure}}_i \gets -\frac{m_i}{\rho_i} \nabla p_i$
        \State Calculate $F^{\text{viscosity}}_i \gets m_i \nu \nabla^2 \mathbf{v}_i$
        \State Calculate $F^{\text{external}}_i$, e.g. $mg$
        \State Calculate $F^{\text{total}}_i = F^\text{pressure}_i + F^\text{viscosity}_i + F^\text{external}_i$
     \EndFor
\EndFor
\ForAll{\text{$n$ particles $i$}}fi
    \State Calculate acceleration, $a_i \gets F_i/m_i$
    \State Calculate new position and velocity (\ref{euler})
\EndFor
\end{algorithmic}
\end{algorithm}

SPH relies on the fact that each particle affects nearby particles (neighbors), so we need to calculate how much each neighbor particle influences the current. We do this by using a kernel function (\ref{kernel}) based on a chosen smoothing distance, $h$. Properties of the SPH fluid at a 3D position can be represented by a smoothed aggregation of values from nearby particles within this smoothing distance. A popular SPH kernel is the cubic-spline function, where $\alpha_d$ is a dimensionality constant to normalize the function \cite{Monaghan1999}.

\begin{equation}
W(r, h) = \alpha_d
\begin{cases}
\frac{2}{3} - \left(\frac{r}{h}\right)^2 + \frac{1}{2}\left(\frac{r}{h}\right)^3, & 0 \le r < h, \\[6pt]
\frac{1}{6}\left(2 - \frac{r}{h}\right)^3, & h \le r < 2h, \\[6pt]
0, & r \ge 2h,
\end{cases}
\label{kernel}
\end{equation}
\vspace{0.3\baselineskip}

In order to retrieve each particle within this smoothing distance, rather than iterating over all $n$ particles for each particle, which has $O(n^2)$ time complexity, we can precalculate each particle's $m$ neighbors using a spatial hash implemented over a uniform grid, which has $O(n\times m)$ time complexity. This enables a real-time simulation with multiple physics substeps, especially since both algorithms are parallelizable and can be implemented on the GPU with compute shaders. 

For our simulation, we use semi-implicit Euler integration (\ref{euler}), which is more stable than the naive Euler method and requires fewer physics substeps than more advanced integration schemes.

\begin{align}
\begin{split}
\mathbf{v}(t + \Delta t) &= \mathbf{v}(t) + \mathbf{a} \Delta t  \\
\mathbf{x}(t + \Delta t) &= \mathbf{x}(t) + \mathbf{v}(t + \Delta t) \Delta t
\label{euler}
\end{split}
\end{align}
\vspace{0.3\baselineskip}

The density of each particle can be calculated by summing the masses of their neighboring particles' weighted by the kernel function (\ref{density}). Although other pressure calculations (e.g. variations of the Tait equation \cite{Dymond1988}) may be more accurate, for stability in real-time we saw that a simple linear pressure calculation (\ref{pressure}) from density $\rho_i$ and rest density $\rho_0$ with a pressure constant $k$ performed best.\\

\begin{equation}
\rho_i = \sum_j m_j \cdot W(|x_i - x_j|,h)
\label{density}
\end{equation}
\vspace{0.3\baselineskip}

\begin{equation}
p_i = k(\rho_i - \rho_0)
\label{pressure}
\end{equation}
\vspace{0.3\baselineskip}

\subsection{Unified Simulation Framework}\label{subsec3.2}

\begin{figure}[b]
    \centering
    \includegraphics[width=0.5\linewidth]{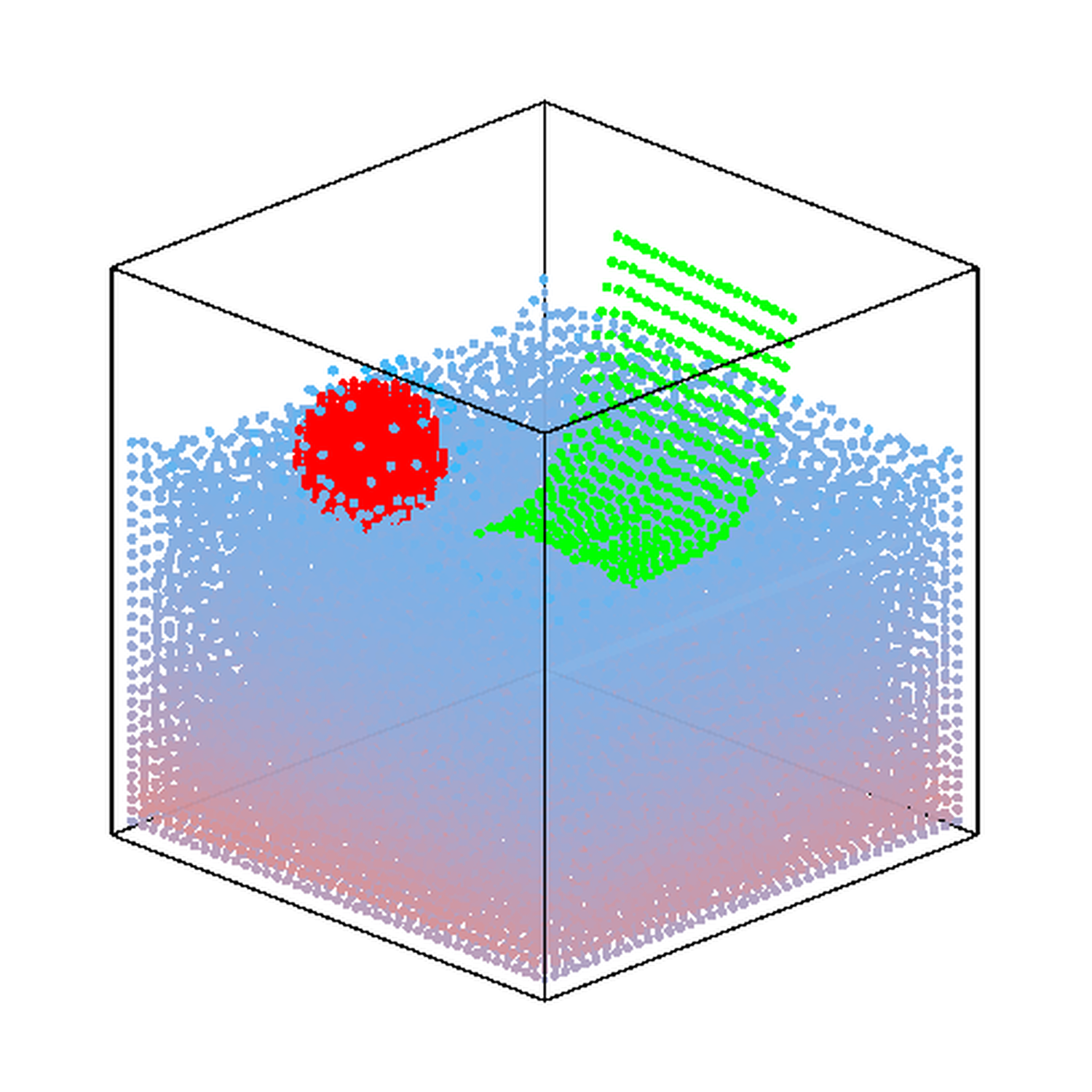}
    \caption{A rigid sphere and soft cloth plane represented as particles within the unified SPH simulation. SPH fluid particles are colored by pressure.}
    \label{fig:particles-bodies}
\end{figure}

In developing our real-time multiphysics simulation framework, one of the primary challenges was achieving stable and efficient fluid--body interaction without sacrificing physical accuracy or simulation speed. This required a method capable of representing both continuous media and discrete objects within a single computational model, while remaining scalable to thousands of interacting elements. To meet these requirements, we adopt a particle-based approach that enabled unified simulation of fluids, soft bodies, and rigid bodies within a single system.

SPH was chosen for this application because it enables us to unify all types of objects as particles within the domain of the SPH simulation as required. In other words, we can model each rigid and soft body as a collection of particles with constraints that enable them to directly interact with the fluid. The volume of each body is discretized and represented as shown in Figure \ref{fig:particles-bodies}. SPH is a common approach for these types of unified simulations \cite{Yan2016, Solenthaler2007, Hashemi2012, Liang2016}.

In order to handle coupled interactions, we give each particle a type and enable/disable integration features for them based on their type. For example, rigid bodies must be tightly packed to repel fluids. To achieve this, we enable them to interact with fluid particles, but integrate them separately. Otherwise, they would disperse within the fluid, violating their rigidity constraints. However, SPH pressure forces are integrated for all other particle interactions.

Rigid particles must maintain a constant relative displacement from their parent object's transformation. In order to maintain this constraint, while still allowing rigid bodies to move properly, we calculate the acceleration of each particle that makes up the rigid body, and apply a point force to the rigid parent. Once the transform of the parent rigid body has been calculated, we can set the relative position of each particle and retrieve their accelerations again. In doing this, the particle positions are passed to and from the CPU, which does have some performance drawbacks.

Soft bodies, in contrast, are still integrated with fluid particle forces, and aptly soft constraints are applied to them. We model soft bodies with a mesh surface, using spring forces that maintain the distance between the particles smoothly over time. We generate a lookup table between two particle indices for each spring, which is calculated in another relatively low-impact compute shader kernel. The full simulation loop is described in Algorithm \ref{alg:sim}.

\begin{algorithm}
\caption{Simulation Loop}\label{alg:sim}
\begin{algorithmic}
\For{\textbf{each} update cycle}
    \For{\textbf{each} physics substep}
        \ForAll{particles} \Comment{Spatial Hash Compute Kernels}
            \State Calculate uniform grid cell
        \EndFor
        \ForAll{grid cells}
            \State Calculate number of particles in cell
        \EndFor
        \ForAll{grid cells}
            \State Calculate cumulative sum of cell particle count
        \EndFor
        \ForAll{particles}
            \State Calculate sorted index using cell cumulative index
        \EndFor \\
        \ForAll{particles} \Comment{Integration Compute Kernels}
            \State Apply soft body forces from springs
        \EndFor
        \ForAll{particles}
            \State Calculate SPH densities and pressures
        \EndFor
        \ForAll{particles}
            \State Integrate SPH forces
        \EndFor
    \EndFor \\
    \ForAll{particles} \Comment{Isosurface Compute Kernels}
        \State Calculate isosurface density field
    \EndFor
    \ForAll{density grid cubes}
        \State Calculate isosurface triangles
    \EndFor \\
    \ForAll{rigid bodies} \Comment{CPU Operations}
        \State Resolve new rigid forces from particles and integrate
    \EndFor
\EndFor
\end{algorithmic}
\end{algorithm}

\subsection{Haptics Integration}\label{subsec3.3}

One of the main problems this research aims to address is the implementation of stable and responsive haptic force feedback within a unified SPH simulation such as the one we propose. The 6-DoF haptic device essentially has one input, a three-dimensional force vector that must be updated every physics cycle in the game engine. In order to calculate this force, rather than estimating the flow of the fluid field at a singular point, which has been a traditional approach, we instead opt to receive the forces from a rigid body.

In the simulation, any rigid body marked as dynamic can be grabbed by pressing a button on the haptic device. In doing this, the haptic device now accepts all of the forces applied to the grabbed point on the rigid, rather than integrating those forces on the rigid within the simulation. Because the rigid body is part of the SPH integration loop, it receives new SPH forces from neighboring particles every update cycle, which are felt by the haptic device. As soon as the user releases the grabbed object, accelerations are once again integrated on the rigid body, and the haptic force is terminated. We can model a variety of different objects using this method, simply converting any closed body into a set of particles that are represented within the rigid framework of the simulation already.

Notably, however, there is some latency between the haptic device and simulation, given the limitation of the game engine update cycle at around 60 Hz. This creates an unstable feedback loop in which the haptic device is forced in one direction, and then consequently the opposite once the rigid body's new position is recognized and likely intersecting with another simulation object. In order to reduce this feedback loop, we choose to take a rolling average of the haptic forces over several update steps, reducing immediate push-backs and recognizing forces for a longer period. Smoothing was done on the order of tens of update cycles, which was found to damp oscillations without reducing responsiveness. The full haptic control system can be seen in Figure \ref{fig:diagram}.

\subsection{Rendering and Visualization}\label{subsec3.4}

Achieving visually realistic rendering of particle-based fluids in real time presents several challenges, as fluids exhibit complex properties such as reflection, refraction, and transparency that depend on a coherent surface definition. To overcome this, we opt to construct an isosurface from a density field generated by the fluid particles. Essentially, a point is enclosed by the isosurface if and only if its density is greater than a defined isovalue.

\begin{figure}[b]
    \centering
    \includegraphics[width=0.5\linewidth]{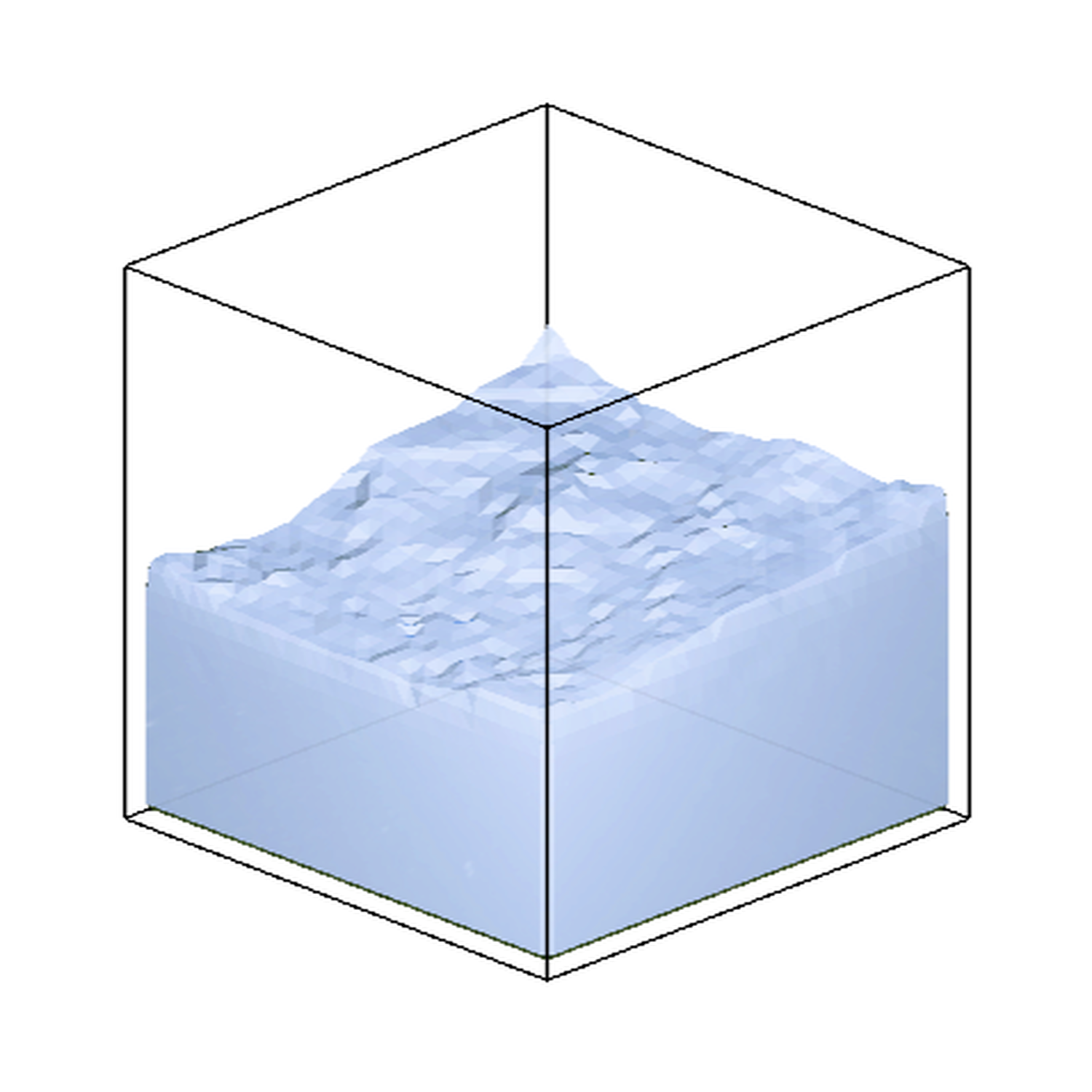}
    \caption{The real-time isosurface generated using the marching cubes algorithm in parallel on the GPU.}
    \label{fig:isosurface}
\end{figure}

To generate the isosurface, the marching cubes algorithm \cite{Lorensen1987} is used because it is parallelizable and allows us to generate triangles on the GPU, enabling the detailed surface to change in real-time. The algorithm works by first establishing a uniform grid of cubes upon which to generate triangles. For each cube corner in the grid, we then calculate the density based on particle proximity and the smoothing kernel. We reuse the uniform grid hash in order to speed up this calculation. To reduce artifacts from aligned grid cells, the marching cubes grid is offset from the uniform grid. Once the density field is calculated, a contour is then estimated between each point and interpolated based on density values. Any point below the isovalue will not be included within the isosurface that represents the fluid volume. Figure \ref{fig:isosurface} shows the resulting isosurface used in our simulation framework rendered in-engine.

Rendering large-scale particle and surface-based simulations in real time within Unity also presents significant challenges in terms of performance, depth precision, and visual realism. To address these issues, we implement a series of shaders that utilize custom rendering techniques.

For particle system rendering, performance was constrained largely by GPU bandwidth when displaying thousands of fluid particles, so we used an instanced rendering approach via Unity's StructuredBuffer object to stream particle colors and positions directly to the GPU. Billboarding was handled within the vertex shader using camera-aligned quads, and linear depth per particle was calculated and used along with a dithering matrix to achieve smooth transparency without the need for complex sorting.

For the cloth visualization, we utilized a similar approach, streaming particle positions directly to the shader and reconstructing the mesh procedurally in the vertex shader. Rigid bodies are much simpler and can be rendered as regular 3D meshes with linked transforms.

To render the surface, the triangles for which are already calculated using the marching cubes algorithm, we draw a procedural mesh using Unity's RenderPrimitives function. The compute shader outputs vertex and normal data directly into a GPU buffer, which is then bound to a surface shader which incorporates cubemap reflections, fresnel-based highlights, and adjustable transparency and specular controls. Together, these Unity specific methods allowed real-time visualization of fluids, soft bodies, and surfaces with reduced CPU overhead and improved visual quality.

\begin{figure}[t]
    \centering
    \includegraphics[width=0.5\linewidth]{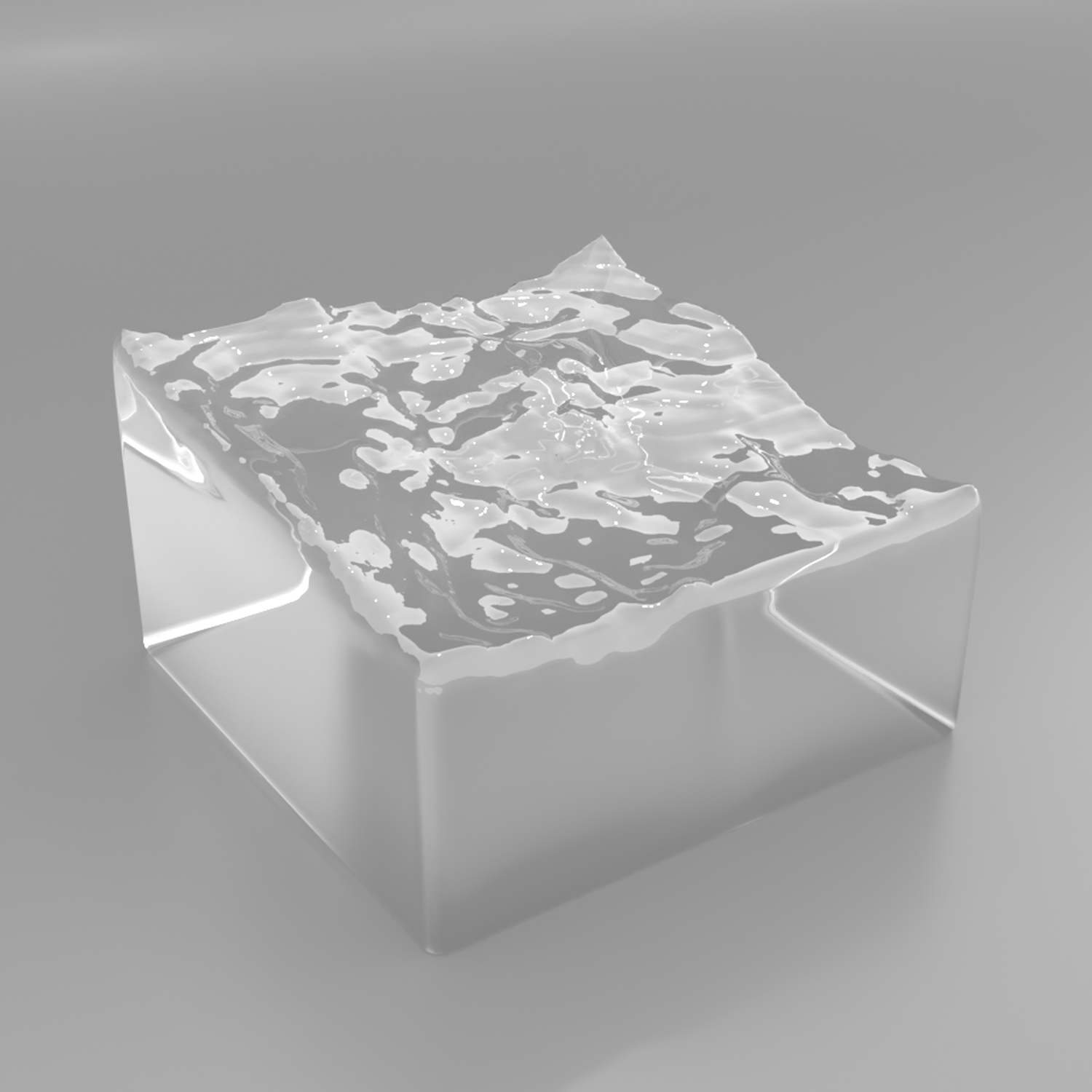}
    \caption{A single ray traced frame of the recorded OBJ fluid sequence from Unity. Rendered in Blender with Cycles utilizing a principled BSDF shader. }
    \label{fig:water}
\end{figure}

Additionally, the fluid isosurface can be recorded as a series of OBJ files which can then be rendered in an external program. For example, we can render a high-resolution ray traced animation of the fluid interactions in Blender, a rendered frame of which can be seen in Figure \ref{fig:water}.

\subsection{Game Engine Integration}\label{subsec3.5}

A major limitation in current physically based simulation research is the lack of modular, easily modifiable tools that can be integrated directly within modern game engines. To address this, an aim of this research was to develop a simulation framework that is extensible and easy to implement into a variety of scenarios. For this project, we used Unity \cite{unity6000}, which is popular among developers for its ease of use. However, our method can be easily extended to other engines that support compute shaders.

We used the Haptics Direct plugin \cite{hapticsdirect} for Unity to integrate the 6-DoF haptic device, and existing virtual reality support for Unity to integrate an HTC Vive Pro 2 headset \cite{vive}. 

In order to enable the creation of educational applications that utilize our simulation, we developed several tools that can be used to construct many different scenarios. One such tool is a component that converts any object with a convex collision component into discrete particles to be used as a rigid body. Rigid bodies can either be dynamic, such as a paintbrush, or static, such as a bowl that holds water. Another tool is a fluid inflow domain that enables the developer to create fluids of varying amounts and densities. We additionally created a tool that can create a soft body cloth plane of any size, with definable anchor points. Finally, many settings are exposed on the actual fluid domain component, which allows the developer to tweak things like viscosity, haptic strength, or even the shader the fluid is rendered with.

The application can be run with any number of rigid bodies, soft bodies, or inflows, and with or without a virtual reality headset. We developed several educational applications including a multi-phase fluid density physics demonstration, a painting tool with color mixing, and a simulated surgery with a scalpel, incision, and blood.

\section{Results}\label{sec4}

To determine the accuracy and practicality of the proposed simulation framework, we analyze both the qualitative experience of user interaction and quantitative performance metrics. In this section, we discuss these aspects of our framework by examining the demonstration applications we designed to test the features and limits of the simulation.

\subsection{Qualitative and Stability analysis}\label{4.1}




\begin{figure*}[t]
\centering
\begin{minipage}{0.31\textwidth}
    \centering
    \includegraphics[width=\linewidth]{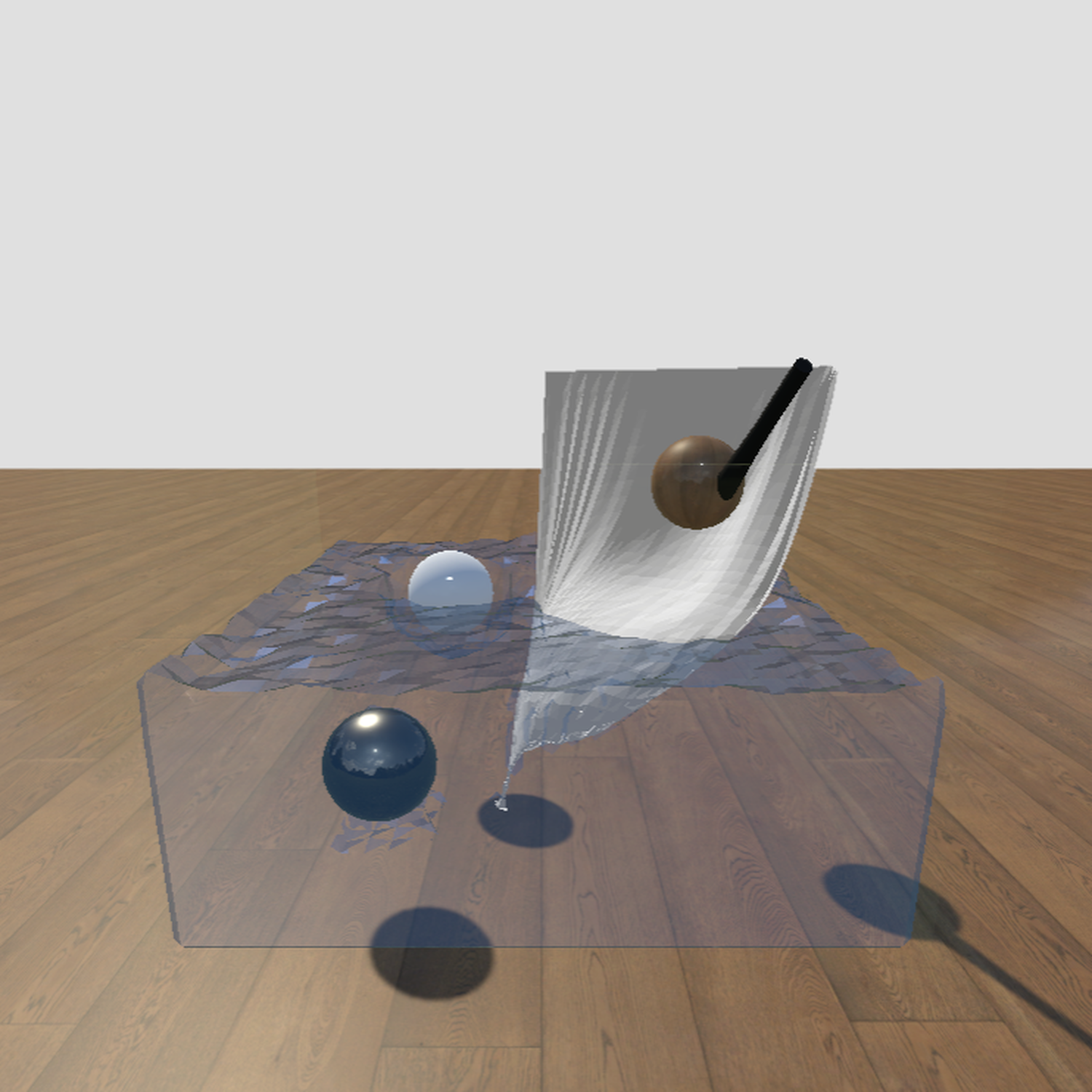}
    \caption{Three balls of different densities are shown interacting with water and a cloth plane and are manipulated using the haptic device. The wood, plastic, and metal balls have densities of $0.1$, $0.3$, and $1.0$, and the water has a density of $1.0$.}
    \label{fig:demo1}
\end{minipage}%
\hfill
\begin{minipage}{0.31\textwidth}
    \centering
    \includegraphics[width=\linewidth]{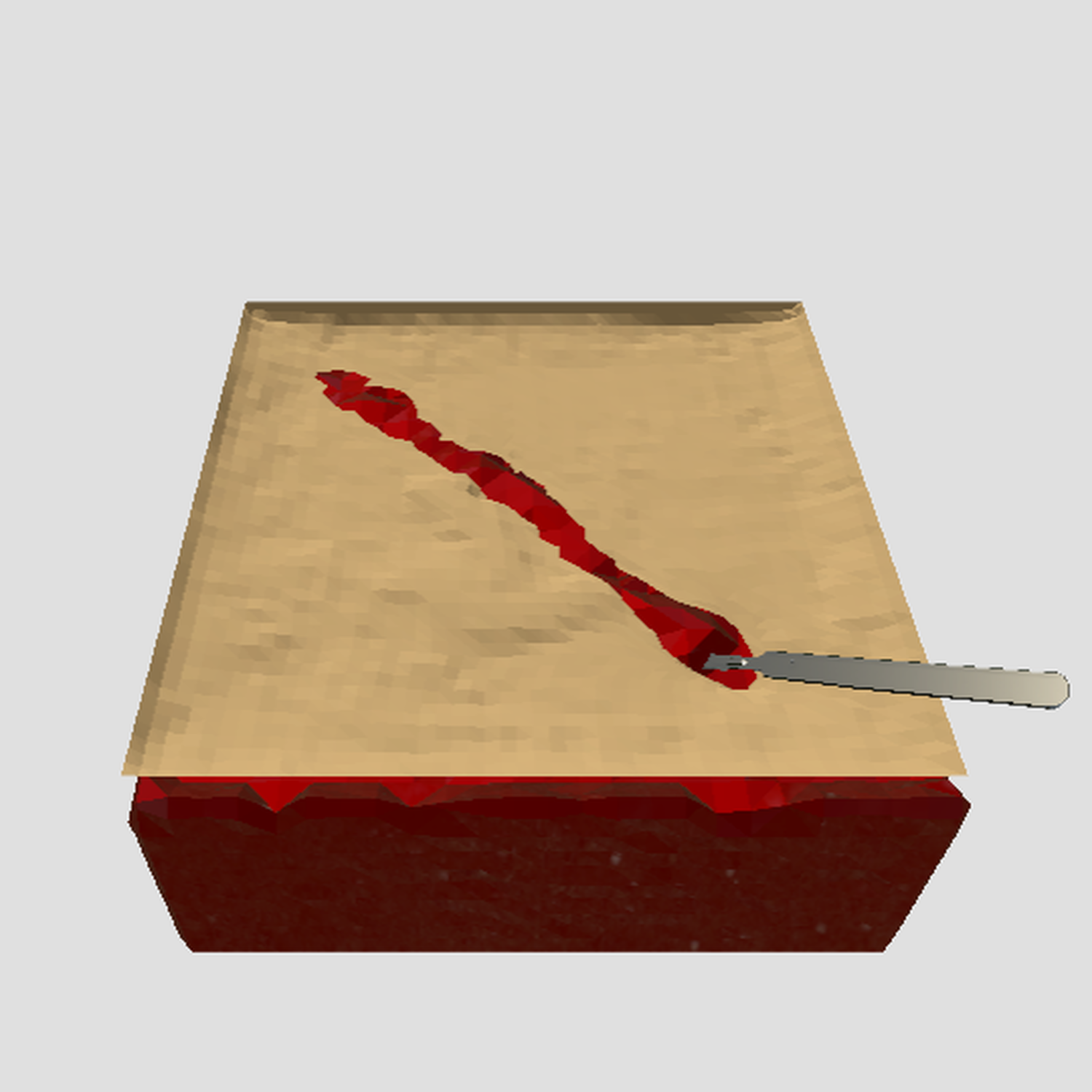}
    \caption{The user interacting with a surgical scene, holding a scalpel and performing an incision on a soft body that represents skin with fluid blood underneath. Force from cutting the skin as well as from the blood can be felt.}
    \label{fig:demo2}
\end{minipage}%
\hfill
\begin{minipage}{0.31\textwidth}
    \centering
    \includegraphics[width=\linewidth]{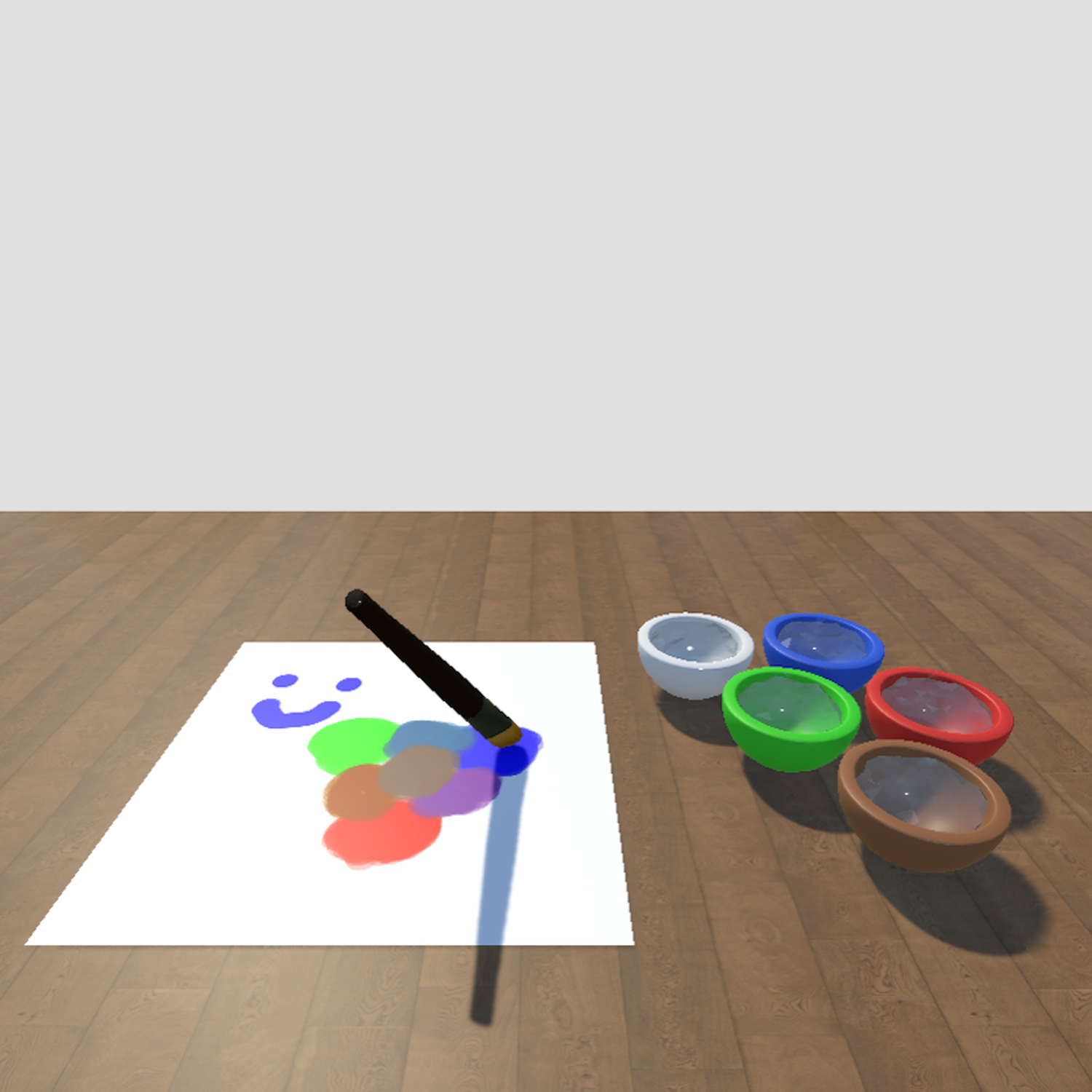}
    \caption{A painting scenario where the user can utilize a paintbrush to mix and paint with various colors. There is not only force feedback from swirling the brush in the bowls, but also from pressing down on the canvas.}
    \label{fig:demo3}
\end{minipage}
\end{figure*}

Figures \ref{fig:demo1}, \ref{fig:demo2}, and \ref{fig:demo3} show the three scenarios we designed in order to showcase the various capabilities of our tools. Within the bounds of these scenarios, the simulation was accurate and maintained stability over time. Given the real-time and hardware constraints, the fluid flow was not perfectly accurate, but it has the potential to be so with further optimizations and better hardware.

Each scenario runs in real-time around 60 Hz. The simulation can be recorded live as a sequence of OBJ files for later rendering at little cost to the stability and performance of the simulation. A summary of performance data for each scenario is given in Table \ref{tab:results}.

\begin{table}[hbt!]
\centering
\caption{Performance metrics for the three demonstration scenarios using the proposed real-time SPH--haptic framework. Each simulation was run on an NVIDIA RTX 3080 Mobile GPU (150W TGP)}
\resizebox{\linewidth}{!}{%
\begin{tabular}{@{}lcccccc@{}}
\toprule
\textbf{Scenario} & \shortstack{\textbf{Particle}\\\textbf{Count}} & \shortstack{\textbf{Physics}\\\textbf{Substeps}} & \shortstack{\textbf{Avg.}\\\textbf{FPS}} & \shortstack{\textbf{Avg. Tri}\\\textbf{Count}} & \shortstack{\textbf{GPU}\\\textbf{Util.}} \\ \midrule
Rigid Balls/Cloth Plane & 45,284 & 4 & 42 & 31,792 & 75\% \\
Surgical Scene & 24,217 & 4 & 89 & 23,669 & 59\% \\
Painting Simulation & 7,734 & 4 & 118 & 4,928 & 37\% \\ \bottomrule
\end{tabular}
\label{tab:results}
} 
\end{table}

\subsection{Performance Analysis}\label{4.2}

\begin{figure}[b]
    \centering
    \includegraphics[width=1.0\linewidth]{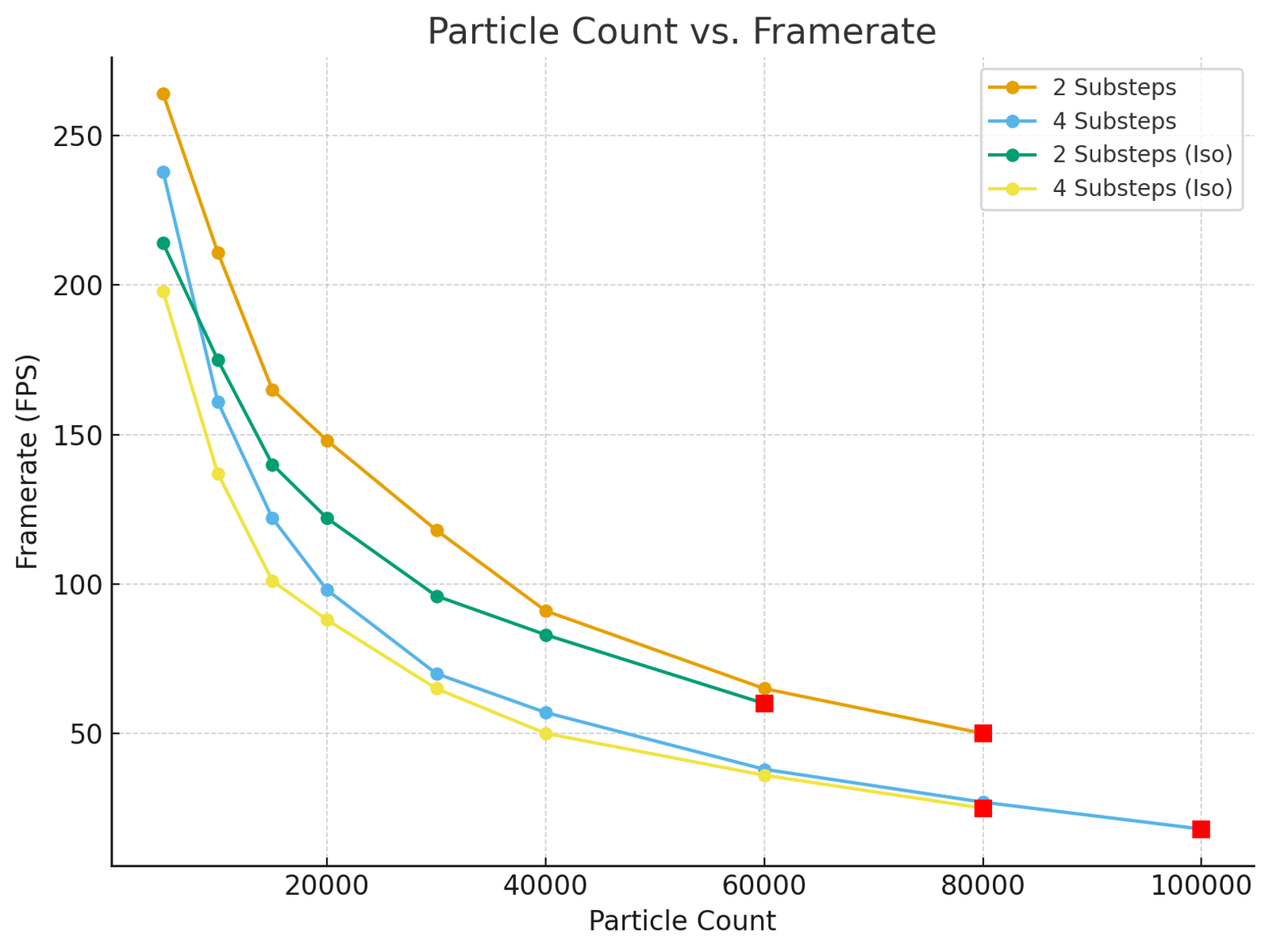}
    \caption{Framerate vs. particle count tested with different simulation settings. Red squares indicate a particle count past which the simulation became unstable unpredictably.}
    \label{fig:results}
\end{figure}

\begin{figure}[b]
    \centering
    \includegraphics[width=0.8\linewidth]{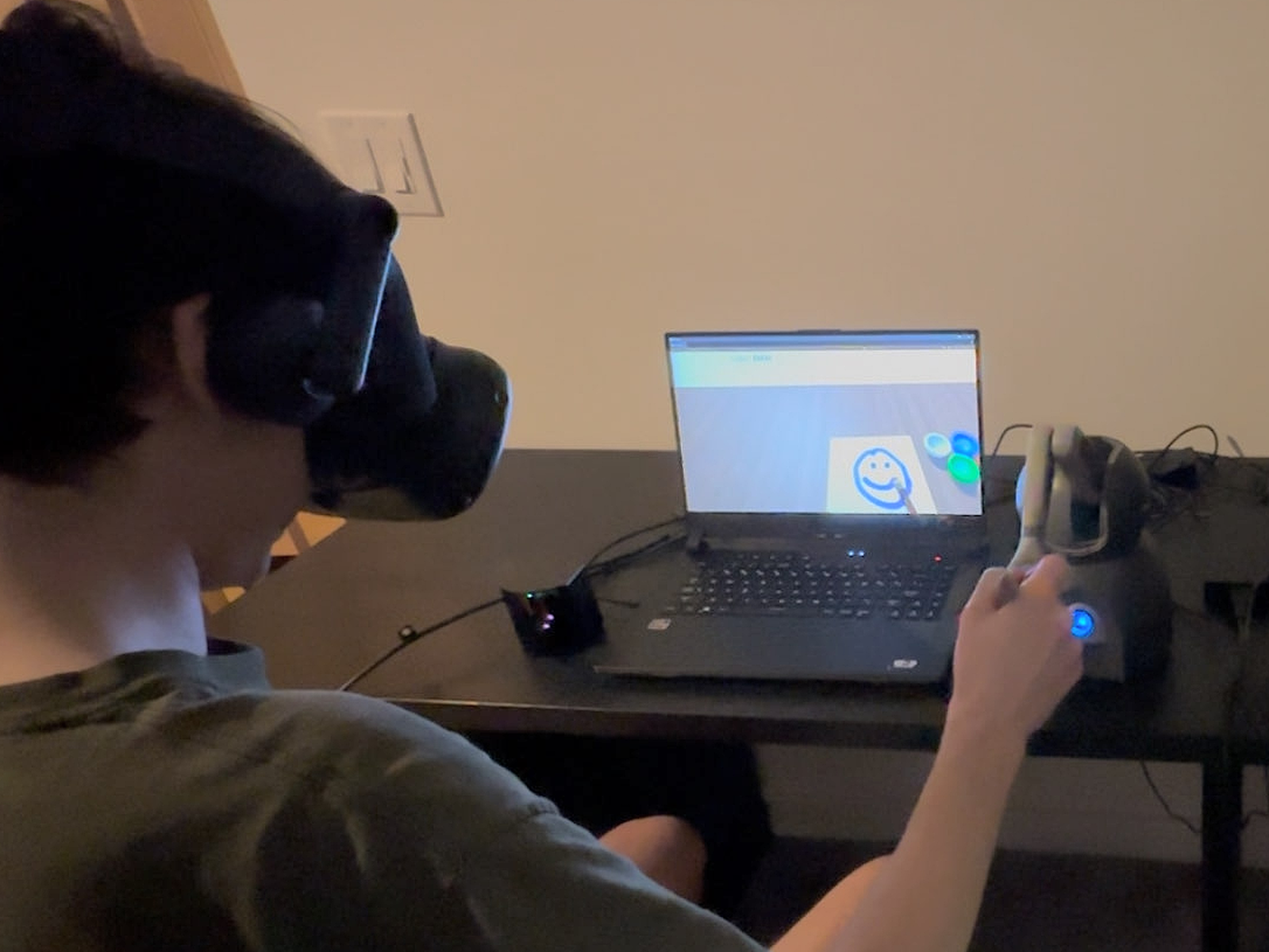}
    \caption{The simulation framework is demonstrated using a laptop featuring an RTX 3080 Mobile GPU, an HTC Vive Pro 2 HMD, and a 3D Systems Touch haptic device to run the painting demo.}
    \label{fig:using}
\end{figure}

While our simulation framework is highly scalable on the GPU, it is important to understand its limits and analyze the performance under various loads on consumer grade hardware. Figure \ref{fig:results} shows the simulation framerate with respect to particle count for various conditions. All data was collected on a consumer grade laptop with an NVIDIA RTX 3080 Mobile GPU using up to 150W total graphics power (TGP). Figure \ref{fig:using} shows the experimental setup used for data acquisition, including the laptop running the simulation, the 6-DoF haptic device, and the VR HMD.

At low particle counts, the simulations without an isosurface run at a higher frequency, but as the particle count is increased to a more reasonable amount, we can see that the simulations with two physics substeps run somewhat faster than the four substep equivalents, which is to be expected. However, for higher particle count runs, more physics substeps are able to achieve a more stable simulation, indicating that while more accurate simulations might not be able to run in real-time on this hardware, increasing the physics substeps achieves a better simulation stability. 

Moreover, our simulation framework performed well for particle counts around 50,000, hitting framerates around the 60 Hz mark depending on the number of physics substeps used, which is satisfactory for real-time interactions.

\section{Future Work}\label{sec5}

Future work on this project could focus on both improving simulation accuracy, as well as overall simulation stability and performance. The force feedback perceived during interaction with the fluid and rigid bodies is generally accurate in direction, but the magnitude is not always well matched to expectations, and tuning the force parameters remains an empirical process. This behavior results from both the simulation's resolution and the implementation of haptic interaction handling. Future work could focus on improving the accuracy of force feedback in real-time simulations such as this one through better handling of force integration under low-granularity data constraints that are necessary for real-time applications.

Additionally, this project, as well as future implementations, could benefit greatly from the optimization of the SPH algorithm. Although spatial hashing with a uniform grid was performant enough to make our simulation real-time on the GPU, future works could investigate the implementation of smarter hashing algorithms that are more friendly to the GPU cache \cite{Ihmsen2010} or even use existing SPH implementations that have already been fine-tuned to be as optimized and as stable as possible.

This work can also be expanded to other applications with the creation of more game engine tools that are modular and easy to drop into existing projects. The tools we propose are already able to create a wide variety of scenarios, but integration with other Unity plugins, add-ons, and existing setups may prove difficult. Having the simulation integrate directly into the physics system of a game engine would be challenging, but reduce the overhead of recreating complicated physics schemes within our framework.

\section{Conclusion}\label{sec6}

This research has demonstrated that a unified SPH simulation is feasible to represent real-time interactions between rigid and soft bodies and fluids with force-feedback. We have created a robust multiphysics simulation framework that is able to cover a wide variety of scenarios and applications and can be used within a popular game engine.

In designing this framework, we employed a novel combination of methods in order to achieve a stable, real-time simulation. Existing SPH algorithms were used and adapted as a way to unify interactions between simulation objects. Parallelizable algorithms were also utilized to accelerate the simulation on the GPU for high-performance iteration and rendering suitable for virtual reality. In addition, we developed an innovative method for handling haptic interactions in such a simulation by receiving forces from a rigid body rather than using a single-point fluid flow estimation. To demonstrate the capabilities of our project, we created multiple example scenarios to show how our simulation framework tools could be used within the Unity game engine.

This research has a broad influence and can be useful for applications in computer graphics, scientific computing, and virtual reality education tools. Handling coupled interactions between objects that behave differently from one another is a challenging and ongoing problem with many solutions, and receiving haptic force-feedback adds yet another layer of complexity on top. Ideas from this project may be reused, extended, and combined with other methods for even more performant and accurate simulations in the future. In creating and evolving such a framework, we can teach about the real world without real consequences, imagine rare scenarios in order to rigorously test resilient systems, and even foster creativity through games and educational tools.

\begin{acks}
This research received a Dean's Undergraduate Research Fund (DURF) grant from the College of Arts and Science (CAS) at New York University (NYU).
\end{acks}

\bibliographystyle{ACM-Reference-Format}
\bibliography{bib}

\end{document}